# Drastic changes in electronic, magnetic, mechanical and bonding properties from $Zr_2CoH_5$ to $Mg_2CoH_5$.

Samir F. Matar[a,b,*]

[a] CNRS, ICMCB, UPR 9048, F-33600 Pessac, France.

[b] Univ. Bordeaux, ICMCB, UPR 9048, F-33600 Pessac, France.

*Mail :* matar@icmcb-bordeaux.cnrs.fr

*Abstract:*

Despite similarities in formulae and local structures of $Zr_2CoH_5$ and $Mg_2CoH_5$, they are shown from ab initio calculations to present contrasted electronic, magnetic, mechanical and bonding properties due to the environment of cobalt with hydrogen characterized by negatively charged $CoH_x$ entities (x = 4, 5 resp.) and to the chemical nature of Zr versus Mg. $Zr_2CoH_5$ is found more cohesive, harder and less ductile than $Mg_2CoH_5$. High density of states at the Fermi level arises from out-of-plane non-bonding Co-$d_z^2$ in $Zr_2CoH_5$ which is found metallic ferromagnet in the ground state, in contrast with non magnetic and insulating $Mg_2CoH_5$.

## 1. Introduction

The potential candidates for hydrogen storage in metals belong to two main families: the intermetallic (binary, ternary and quaternary) hydrides and the transition metal hydrido-complexes; also there are the saline metal hydrides based on alkaline and alkaline earth elements [1-3]. Several experimental and theoretical reports were devoted to better understand their physical and chemical properties (cf. [4] and refs. cited therein). One initial impetus for such extended research is the non usability of archetype $MgH_2$, characterized by a large hydrogen capacity (7.6 wt.–%), owing to the elevated desorption temperature (~350°C) caused by highly ionic Mg-H bond. Other Mg based compounds were then studied as candidates for hydrogen storage such as $Mg_2CoH_5$ [5] synthesized starting from a mixture of the constituents because $Mg_2Co$ intermetallic does not exist. On the contrary $Zr_2CoH_5$ with similar formulation is an insertion compound obtained by hydrogenation of $Zr_2Co$ intermetallic [6]. While $Mg_2CoH_5$ is a hydrido complex of cobalt, $Zr_2CoH_5$ is an intermetallic hydride. Some structural similarities exist between $Zr_2CoH_5$ and $Mg_2CoH_5$ especially for the Co-H short distances (~1.51 – 1.66 Å) leading to $CoH_x$ (x = 4, 5 resp.) staggered motifs but they are opposed with large differences in the electronic, magnetic, mechanical and bonding properties characterizing them, as shown throughout the paper. On the theoretical side $Mg_2CoH_5$ was calculated [7] as insulating with a small gap and characterized by a low magnitude bulk modulus $B_0$ = 54 GPa.

For a better understanding of the interplay *extended and local structure − property* relationships and the physical and chemical roles played by hydrogen in the title compounds, an original comparative study is presented based on computations within the quantum theoretical density functional DFT framework [8, 9].

## 2. Relevant structural details

The tetragonal crystal structures of the two hydrogenated compounds are described in Table 1 and sketched in Fig. 1. $Zr_2CoH_5$ structure [6] derives from the tetragonal $Zr_2Co$ structure (*I4/mcm* space group SG). A change of SG arises from the occupation of H2 atoms of half of the 32 *Zr₃Co* tetrahedral holes (i.e. 16 atoms) of the intermetallic in an ordered manner. The structure is then described in the *P4/ncc* SG with 4 FU/cell: $Zr_8Co_4H_{20}$. The remaining four hydrogen atoms (H1) are inserted into $Zr_4$ holes which form files of edge sharing tetrahedra along tetragonal *c* axis. The feature of flattened square pyramidal *CoH2$_4$* coordination highlighted in Fig. 1 arises from the shortest interatomic distances in the structure: d(Co-H) = 1.66 Å. Such range of short Co-H distances is found in the cobalt hydrido complex $Mg_2CoH_5$ [5] (Table 1, Fig. 1) which shows similarities with the Zr compound for cobalt surrounding with hydrogen. Four H2 form the square base of the pyramid around cobalt and additional apical H1 completes the pyramid which is characterized by two distances: $d_{Co-H2}$ ~1.51 Å and $d_{Co-H1}$ = 1.59 Å. H1 and H2 are in *CoMg$_4$* environments with $d_{Mg2-H1}$ = 2.23 Å and $d_{Mg2-H2}$ = 2.40 Å, larger than $d_{Co-H1}$ = 1.59 Å and $d_{Co-H2}$ = 1.52 Å. From these observations one may expect differences in bonding due to the coordination polyhedra on one hand and to the nature of Mg ($3s^2$) versus Zr ($5s^2 4d^2$) which have otherwise close electronegativity magnitudes: χ(Mg) = 1.31 and χ(Zr) = 1.33 compared to the more electronegative cobalt (χ = 1.88) on the other hand.

## 3. Computation methods

Two DFT computational methods were used in a complementary manner. The Vienna ab initio simulation package (VASP) code [10, 11] allows geometry optimization and energy calculations. For this we use the projector augmented wave (PAW) method [11, 12], built within the generalized gradient approximation (GGA) scheme following Perdew, Burke and Ernzerhof (PBE) [13]. Also preliminary calculations with local density approximation LDA [14] led to underestimated volumes versus the experiment. The conjugate-gradient algorithm [15] is used in this computational scheme to relax the atoms. The tetrahedron method with Blöchl corrections [12] as well as a Methfessel-Paxton [16] scheme were applied for both geometry relaxation and total energy calculations. Brillouin-zone (BZ) integrals were approximated using the special k-point sampling. The optimization of the structural parameters was performed until the forces on the atoms were less than 0.02 eV/Å and all stress components less than 0.003 eV/Å$^3$. The calculations are converged at an energy cut-off of 400 eV for the plane-wave basis set with respect to the k-point integration with a starting mesh of 4×4×4 up to 8×8×8 for best convergence and relaxation to zero strains. The calculations are scalar relativistic.

Then all-electron calculations with GGA were carried out for a full description of the electronic structure and the properties of chemical bonding, using full potential scalar-relativistic augmented spherical wave (ASW) method [17, 18]. In the minimal ASW basis set, we chose the outermost shells to represent the valence states and the matrix elements were constructed using partial waves up to $l_{max}+1 = 3$ for Zr and Co and $l_{max}+1 = 1$ for H. Self-consistency was achieved when charge transfers and energy changes between two successive cycles were below $10^{-8}$ and $10^{-6}$ eV, respectively. BZ integrations were performed using the linear tetrahedron method within the irreducible wedge. In order to optimize the basis set, additional augmented spherical waves are placed at carefully selected interstitial sites (IS). Besides the site projected density of states, we discuss qualitatively the pair interactions based on the overlap population analysis with the crystal orbital overlap population (COOP) [19]. In the plots, positive, negative, and zero COOP indicate bonding, anti-bonding, and non-bonding interactions, respectively.

**4. Results from PAW-GGA calculations.**

*4.1 Geometry optimization and cohesive energies*

The experimental and calculated structure parameters are given in Table 1. The latter show some differences with respect to experiment especially for the H2 coordinate which was obtained for the deuteride in both compounds [5, 6]. Nevertheless the agreement is such that it can allow establishing trends between the two compounds as for the energies, the mechanical properties as well as for the relative charge transfers.

The cohesive energies are obtained from energy differences between the total energy of the compound and those of the atomic constituents Zr, Mg and Co as well as dihydrogen which have the following energies from PAW-GGA calculations (in eV units /atom):

Mg (-1.506); Co (-6.888), Zr (-8.477) and 2H (-6.577).

From the total electronic energies given in Table 1, the resulting cohesive energies in eV/atom are:

$E_{coh.}(Zr_2CoH_5)$ = -0.488 eV and $E_{coh.}(Mg_2CoH_5)$ = -0.343 eV.

A larger cohesive energy characterizing $Zr_2CoH_5$ versus $Mg_2CoH_5$ could arise from different bonding interactions within the structure between the metallic constituents on one hand and between hydrogen and the metal substructures on the other hand (section 5).

The directional bonding within the structure involves local structure strains which can be quantified from the calculation of the elastic constants $C_{ij}$ determined by performing finite distortions of the structure lattice.

*4.2 Elastic constants, bulk and shear modules*

In tetragonal symmetry there are six independent elastic stiffness constants $C_{11}$, $C_{33}$, $C_{44}$, $C_{66}$, $C_{12}$ and $C_{13}$. Most encountered compounds are polycrystalline where monocrystalline grains are randomly oriented so that on a large scale, such materials can be considered as statistically isotropic. They are then completely described by the bulk modulus B and the shear modulus G, which may be obtained by averaging the single-crystal elastic constants. The most widely used averaging method of the elastic stiffness constants is Voigt's based on a uniform strain (cf. [20] for a review). The calculated set of elastic constants in $Zr_2CoH_5$ and $Mg_2CoH_5$ are (in units of GPa):

$Zr_2CoH_5$: $C_{11} = C_{22} = 237$; $C_{12} = 107$; $C_{13} = 122$; $C_{33} = 227$; $C_{44} = 57$ and $C_{66} = 83$.

$Mg_2CoH_5$: $C_{11} = C_{22} = 140$; $C_{12} = 20$; $C_{13} = 26$; $C_{33} = 164$; $C_{44} = 59$ and $C_{66} = 55$.

All $C_{ij}$ are positive and their combinations: $C_{11} > C_{12}$, $C_{11}C_{33} > C_{13}^2$ and $(C_{11}+C_{12})C_{33} > 2C_{13}^2$ obey the rules pertaining to the mechanical stability of the two compounds. Nevertheless the values of $Zr_2CoH_5$ are systematically higher which suggests different mechanical properties versus $Mg_2CoH_5$. This is obtained from the calculations of the respective bulk and shear modules, $B_V$ and $G_V$ following Voigt and formulated as follows:

$B_V = 1/9 \{2(C_{11} + C_{12}) + 4C_{13} + C_{33}\}$ and
$G_V = 1/30 \{12C_{44} + 12C_{66} + C_{11} + C_{12} + 2C_{33} - 4C_{13}\}$

The numerical values are then:

$Zr_2CoH_5$: $B_V = 156$ GPa, $G_V = 66$ GPa. $G_V/B_V = 0.42$

$Mg_2CoH_5$: $B_V = 65$ GPa, $G_V = 10$ GPa. $G_V/B_V = 0.15$.

The results show a much harder Zr based compound versus Mg based one. The calculated value of $Mg_2CoH_5$ is close to the one computed with a different method by Zhang et al. [7]. The large difference between the two values can arise from the bulk modules of the constituents in GPa units: B(Zr) = 91, B(Mg) = 45, i.e. with twice less hard Mg, whereas cobalt which is common to the two compounds has a hardness of 180 GPa. The averaging which gives 121 and 90 GPa respectively, does not provide the calculated proportion of Bv values, and other effects such as bonding and chemical nature of the elements come into play. The shear modules which define the rigidity of the material show a significantly lower value for $Mg_2CoH_5$. The corresponding $G_V/B_V$ ratio which is an indicator of brittleness/ductility, has a value of 0.15, almost one third of $G_V/B_V(Zr_2CoH_5)$. Although both ratios are close to ductile transition metals with G/B ratios in the range {0.4 - 0.2} [21], it can be suggested that $Zr_2CoH_5$ is much less ductile (more brittle) than $Mg_2CoH_5$. This further confirms the role of hydrogen bonding.

*4.3 Charge analysis*

We further assess these results by analyzing the charge density issued from the self consistent calculations using the AIM (atoms in molecules theory) approach [22] developed by Bader who devised an intuitive way of splitting molecules into atoms as based purely on the electronic charge density. Typically in chemical systems, the charge density reaches a minimum between atoms and this is a natural region to separate them from each other. Such an analysis can be useful when trends between similar compounds are examined; they do not constitute a tool for evaluating absolute ionizations. Bader's analysis is done using a fast algorithm operating on a charge density grid. The program reads in charge densities obtained from high precision VASP calculations and outputs the total charge associated with each atom. Further in order to include core electrons within the PAW method, use is made of the LAECHG = .TRUE. parameter in the 'INCAR' control file followed by operating a summation of core and valence charge densities using a simple *perl* program [23]. The results of computed charge changes ($\Delta Q$) are such that they lead to neutrality when the respective multiplicities are accounted for; the obtained values are:

$Zr_2CoH_5$: $\Delta Q(Zr) = +1$; $\Delta Q(Co) = -0.373$; $\Delta Q(H1) = -0.433$; $\Delta Q(H2) = -0.318$

$Mg_2CoH_5$: $\Delta Q(Mg) = +0.933$; $\Delta Q(Co) = -0.550$; $\Delta Q(H1) = -0.316$; $\Delta Q(H2) = -0.250$.

Due to the large electronegativity difference between Zr and Mg on one hand and Co on the other hand, charge transfer is observed towards cobalt as well as hydrogen with a larger value from Zr than from Mg. $\Delta Q(Co)$ magnitudes show large differences between the two compounds. In $Mg_2CoH_5$, covalent charges on $H1^{-0.316}$ and $H2^{-0.250}$ are due to the complexation of the five H within the *CoH1$_1$H2$_4$* pyramid characterized by very short distances: $d_{Co-H2}$ ~1.51 Å and $d_{Co-H1}$ = 1.59 Å; also the Co-H covalent bonding is reinforced by large Mg-H distances: $d_{Mg2-H1}$ = 2.23 Å and $d_{Mg2-H2}$ = 2.40 Å. In contrast $Zr_2CoH_5$ shows less covalent behavior for the two different environments for hydrogen: H1 enclosed in the $Zr_4$ tetrahedra with d(Zr-H1) = 2.11 Å has a charge $H1^{-0.433}$, higher than H2 ($H2^{-0.318}$) forming a flattened pyramid with Co (*CoH2₄*) characterized by larger $d_{Co-H2}$ ~1.66 Å. Also H2 is enclosed into *Zr₃Co* tetrahedra and binds with both metals (*vide infra*).

**5. Electronic density of states and chemical bonding properties**.

In as far as fairly good agreements between experimental and calculated crystal parameters are found (Table 1), the electronic density of states and the chemical bonding were analyzed for the two compounds based on experimental data [5, 6] and assuming firstly spin degenerate total spin (NSP) configuration. At self consistent convergence the charge transfer follows the trends observed above with additional charge residues of ~ 0.15 electrons from the atomic spheres to IS.

*5.1 Non magnetic and spin polarized calculations*

Fig. 3 shows the site projected density of states (PDOS) for $Zr_2CoH_5$ and $Mg_2CoH_5$ accounting for site multiplicities and total number of FU per cell. Along the *x* axis the energy is with respect to the Fermi level $E_F$.

In $Zr_2CoH_5$, as expected from the electronic configurations: Zr ($5s^24d^2$) and Co ($4s^23d^7$), i.e. respectively with rather empty and largely filled *d* bands, the valence band (VB) is dominated by Co 3*d* states centered below but close to $E_F$ and Zr 4*d* states centered above $E_F$. The localization of the Zr and Co PDOS is noted with a separation between the VB lower energy part and the *d* PDOS from -3 up to $E_F$. The PDOS structure from -10 to -4 eV arises from hydrogen resulting in similar shapes of Zr/Co and H PDOS, whence the metal-H bonding. Note the sharp Co PDOS at $E_F$ with larger intensity than Zr, especially in view of the twice larger Zr content than Co in the unit cell (Table 1). On the contrary $Mg_2CoH_5$ is semi conducting with a small gap (~0.2 eV) at the top of the valence band (VB), $E_V$. Cobalt *d* states dominate the VB and their itinerant (delocalized) parts from -8 to -3 eV show similar shape with hydrogen, especially H2 which forms the square planar base of the pyramid (Fig. 1).

*5.2 Spin polarized magnetic calculations for $Zr_2CoH_5$.*

In order to assign a role to the Co PDOS at $E_F$ the $d$ states were decomposed into their components, leading to non bonding $d_z^2$ orbital. In fact high PDOS at $E_F$ signal instability in total spins (NSP) configuration within the Stoner mean field theory of band ferromagnetism [24]: The total energy of the spin system stems from the exchange and kinetic energies. Referring the total energy to the non-magnetic state (NSP), this is expressed as:

$$E = constant \{1 - In(E_F)\}.$$

In this expression, I (eV) is the Stoner integral and $n(E_F)$ (1/eV) is the PDOS value for a given state –d– at $E_F$ in the non-magnetic state. If the unit-less Stoner product $In(E_F)$ is larger than 1, energy is lowered and the spin system stabilizes in a (ferrro)magnetically ordered configuration. Then the product $In(E_F)$ provides a stability criterion. From quantum theoretical calculations by Janak [25] a value of I{Co(3d)} = 0.49 eV was derived. With $n(E_F)$ = 2.79 eV$^{-1}$, $In(E_F)$ = 1.37 is calculated and the resulting negative Stoner criterion 1–$In(E_F)$ = -0.37 leads to energy lowering upon the onset of magnetization.

Spin polarized ferromagnetic (SPF) calculations actually lead to a lowering of the total energy by ΔE(SPF-NSP) = -0.11 eV/cell. The total magnetization of M = 0.695 $\mu_B$/FU arises mainly from cobalt with $M_{Co}$ = 0.619 $\mu_B$/FU while $M_{Zr}$ = 0.035 $\mu_B$/FU and vanishingly small moments identified for H and IS. These results are illustrated in Fig. 3 for the site and spin projected PDOS. The magnetic moments arise from a rigid band -like shift between majority ⇑ spin populations and minority ⇓ spins ones through magnetic exchange. This is observed mainly for cobalt while the other constituents PDOS exhibit small (Zr) or no (H1, H2) energy shift.

In order to confirm that the actual magnetic ground state is ferromagnetic, antiferromagnetic SPAF configuration was tested with dispatching the crystal into two magnetic half-subcells, one of them for UP SPINS and the other for DOWN SPINS. The result is zero magnetization as expected from full compensation between the two magnetic subcells, and a raise of the energy by ΔE(SPAF-SPF) = 0.42 eV/cell. Thus the magnetic ground state is predicted to be ferromagnetic.

*5.3 Bonding properties*

Figs. 4 and 5 show the chemical bonding discussed using the COOP criterion based on overlap integrals. The plots account for site multiplicities within each structure, i.e. twice more in $Zr_2CoH_5$ versus $Mg_2CoH_5$. The metal-metal bonding (Figs. 4 a,b) is predominantly hetero atomic, resp. Zr-Co and Mg-Co (Mg1-Co and Mg2-Co). Its magnitude is much larger in the Zr compound in spite of the presence of antibonding states, than in the Mg compound. This arises from the fact that $Zr_2Co$ intermetallic exists whereas $Mg_2Co$ does not. Then the larger cohesive energy of $Zr_2CoH_5$ can be at least partly assigned to this result.

Figs. 5 a,b illustrate the metal-H bonding. In $Zr_2CoH_5$ Zr as well as Co are involved with H1 and H2 bonding. There are three main COOP intensities: Co-H2, Zr-H1 and Zr-H2. They follow respectively from the *CoH2₄* pyramidal motifs, the presence of H1 within *Zr4* tetrahedra and H2 into *Zr3Co* tetrahedra. Nevertheless antibonding COOP with negative intensities are observed especially for Zr-H2 because of the larger contribution of H2 states to the bonding with Co in a competitive manner. Also Co-H1 antibonding states arise from the competitive positive Zr-H1 bonding COOPs. Smaller contribution COOP are found in the neighborhood of $E_F$ with bonding Zr-H2 and antibonding Co-H2, i.e. for the non-bonding Co *d* orbitals responsible for the onset of magnetic moment.

Panel b) is characterized by less pronounced COOP features in $Mg_2CoH_5$, except for the Co-H2 large intensity COOP, i.e. for the bonding of Co with the four H2 forming the square planar basis of the pyramid (Fig. 1). Smaller COOPs are observed for Co-H1 due to the larger Co-H1 separation (1.59 versus 1.51 Å) as well as to the four times less H1 versus H2 (Table 1). In spite of the presence of less anti-bonding COOP the metal − hydrogen bonding is qualitatively larger in $Zr_2CoH_5$. This further assesses its larger cohesive energy

## 6. Conclusion

The presented results have shown differences pertaining to more cohesive, harder and less ductile $Zr_2CoH_5$ versus $Mg_2CoH_5$ caused by larger bonding intensity for inter-metal and metal-hydrogen. A ferromagnetic ground state is identified for the Zr homologue owing to magnetic polarization of out-of-plane non bonding Co-$d_z^2$ versus non magnetic ground state of $Mg_2CoH_5$ owing to $\{CoH_5\}^{4-}$ complex anion which obeys the 18 electrons rule (2) leading to the insulating behavior.

# 7. Acknowledgements:

Part of the computations was done on MCIA-University Bordeaux 1 (UB1) cluster computers. I also thank Mme Catherine Marc of UB1 Main Library.

Table 1: Experimental and (calculated) crystal data for $Zr_2CoH_5$ and $Mg_2CoH_5$.

$Zr_2CoH_5$
SG130 *P4/ncc* Orig. 2
Exp. Ref. [7]

$a$ = 6.926 (6.95) Å
$c$ = 5.647 (5.60) Å
V = 270.88; (270.5) Å$^3$.

| At.(Wyck.) | $x$ | $y$ | $z$ |
|---|---|---|---|
| Co (4*c*) | ¼ | ¼ | 0.0062 (0.005) |
| Zr (8*f*) | 0.4115 (0.411) | 0.5885 (0.589) | ¼ |
| H1 (4*b*) | ¾ | ¼ | 0 |
| H2 (16*g*) | 0.0330 (0.036) | 0.1641 (0.167) | 0.0756 (0.074) |

Total energy (eV)/cell: -176.754 eV (NSP); -177.160 eV (SPF)

$Mg_2CoH_5$
SG129 *P4/nmm* Orig. 2
Exp. Ref. [5]

$a$ = 4.463 (4.40) Å
$c$ = 6.593 (6.57) Å
V = 131.32; (127.31) Å$^3$.

| At.(Wyck.) | $x$ | $y$ | $z$ |
|---|---|---|---|
| Co (2*c*) | ¼ | ¼ | 0.2561 (0.257) |
| Mg1 (2*a*) | ¾ | ¼ | 0 |
| Mg2 (2*b*) | ¾ | ¼ | ½ |
| H1 (2*c*) | ¼ | ¼ | 0.4972 (0.498) |
| H2 (8*j*) | 0.4879 (0.498) | $x$ | 0.2257 (0.229) |

Total energy (eV)/cell: -58.282 eV.

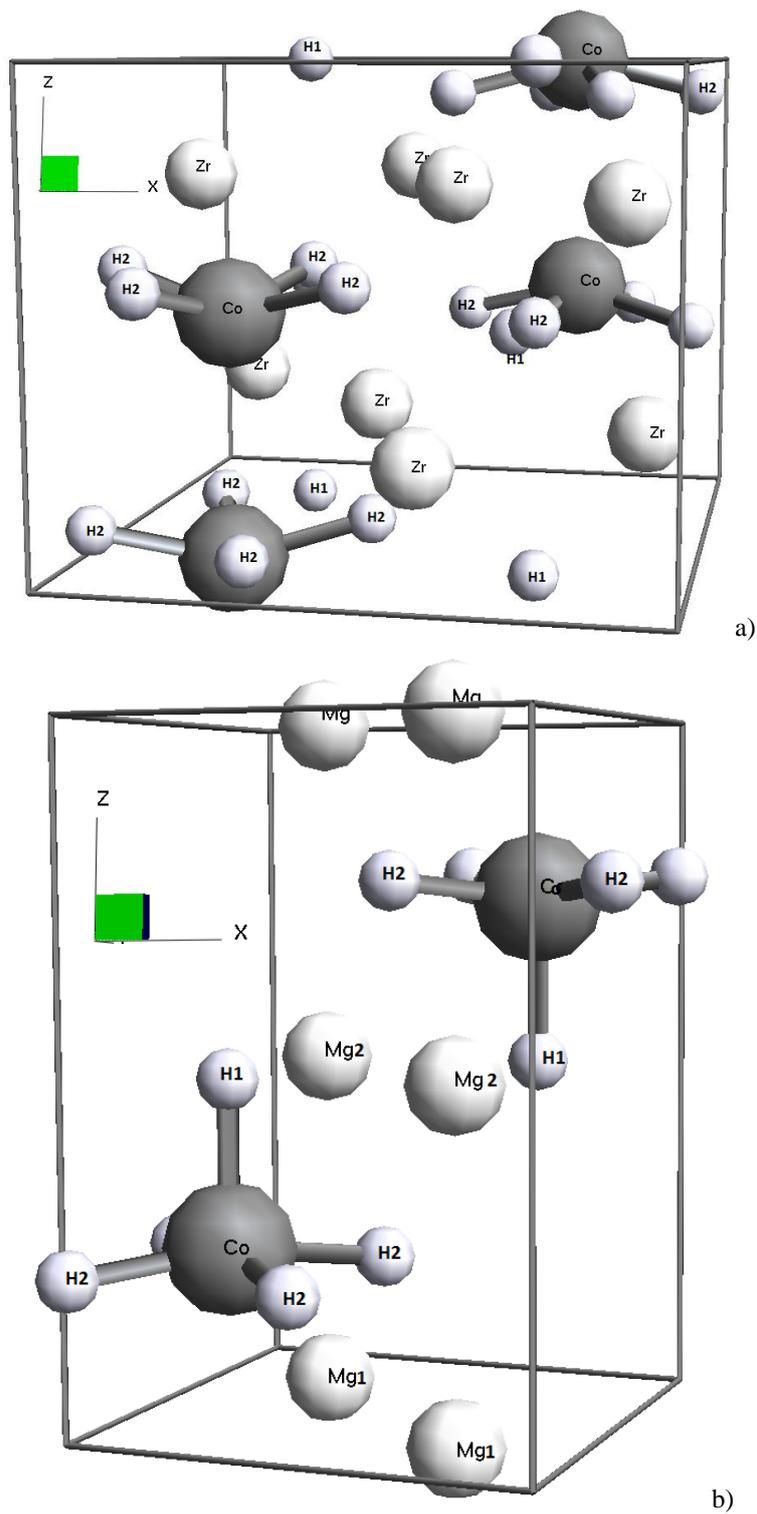

Fig. 1 Perspective views of the crystal structures of a) $Zr_2CoH_5$ highlighting the $\{CoH_4\}$ assemblies and b) $Mg_2CoH_5$ with the *$CoH_5$* square pyramidal complexes. Atom labels are as in Table 1.

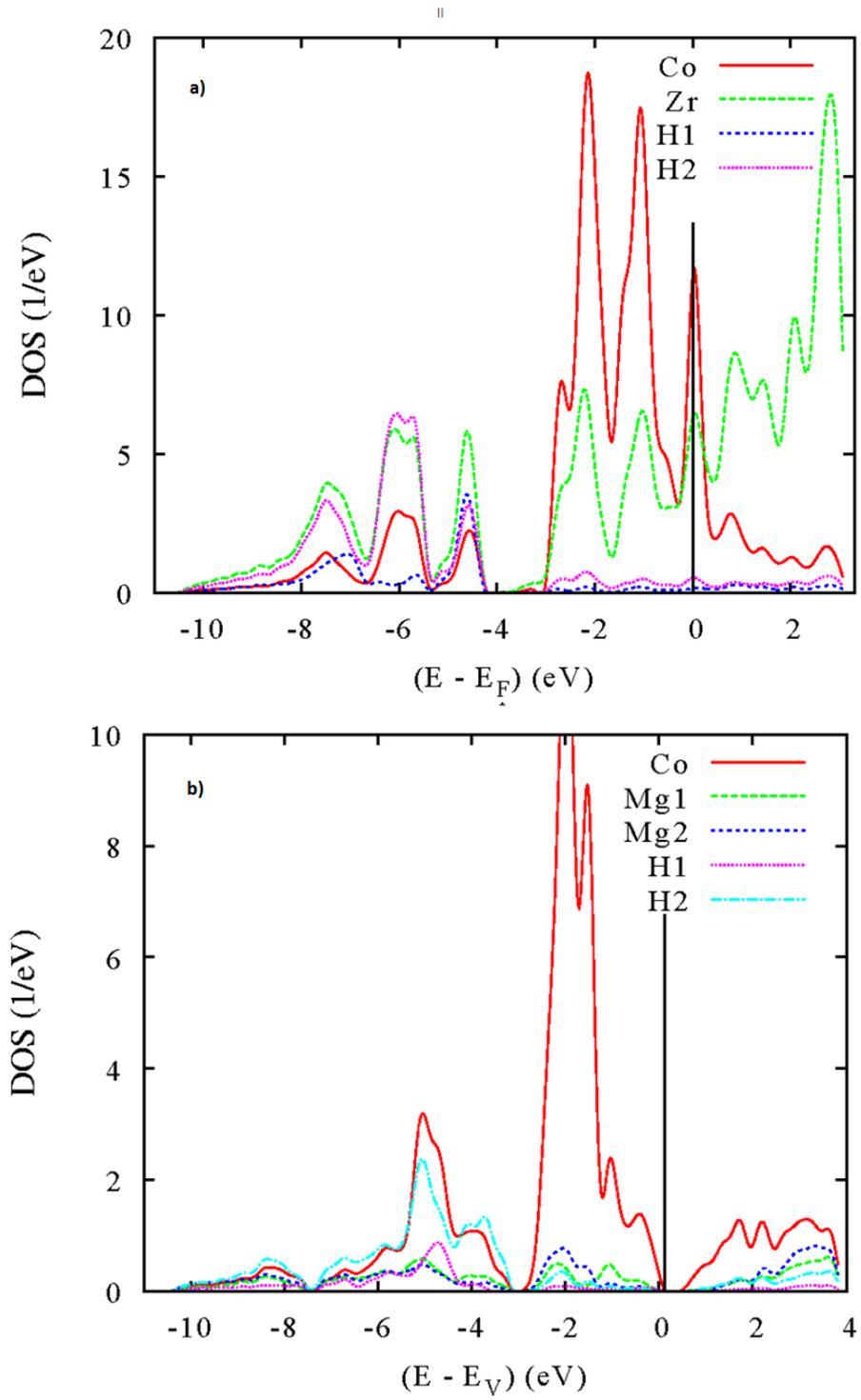

Fig. 2: Site projected density of states (PDOS) for a) $Zr_2CoH_5$ and b) $Mg_2CoH_5$.

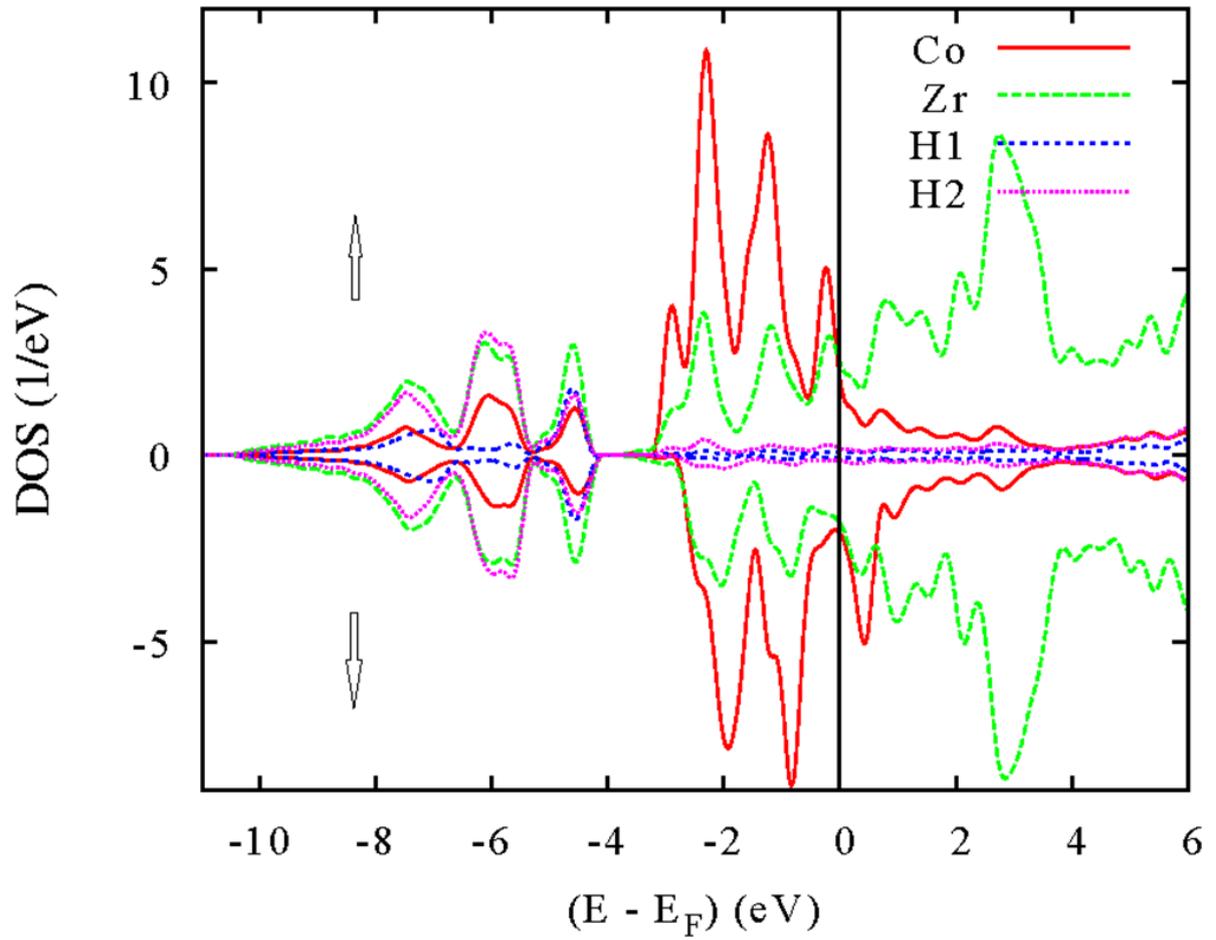

Fig. 3: Spin polarized ferromagnetic $Zr_2CoH_5$: Site projected density of states (PDOS) for the majority (⇑) and minority (⇓) spins.

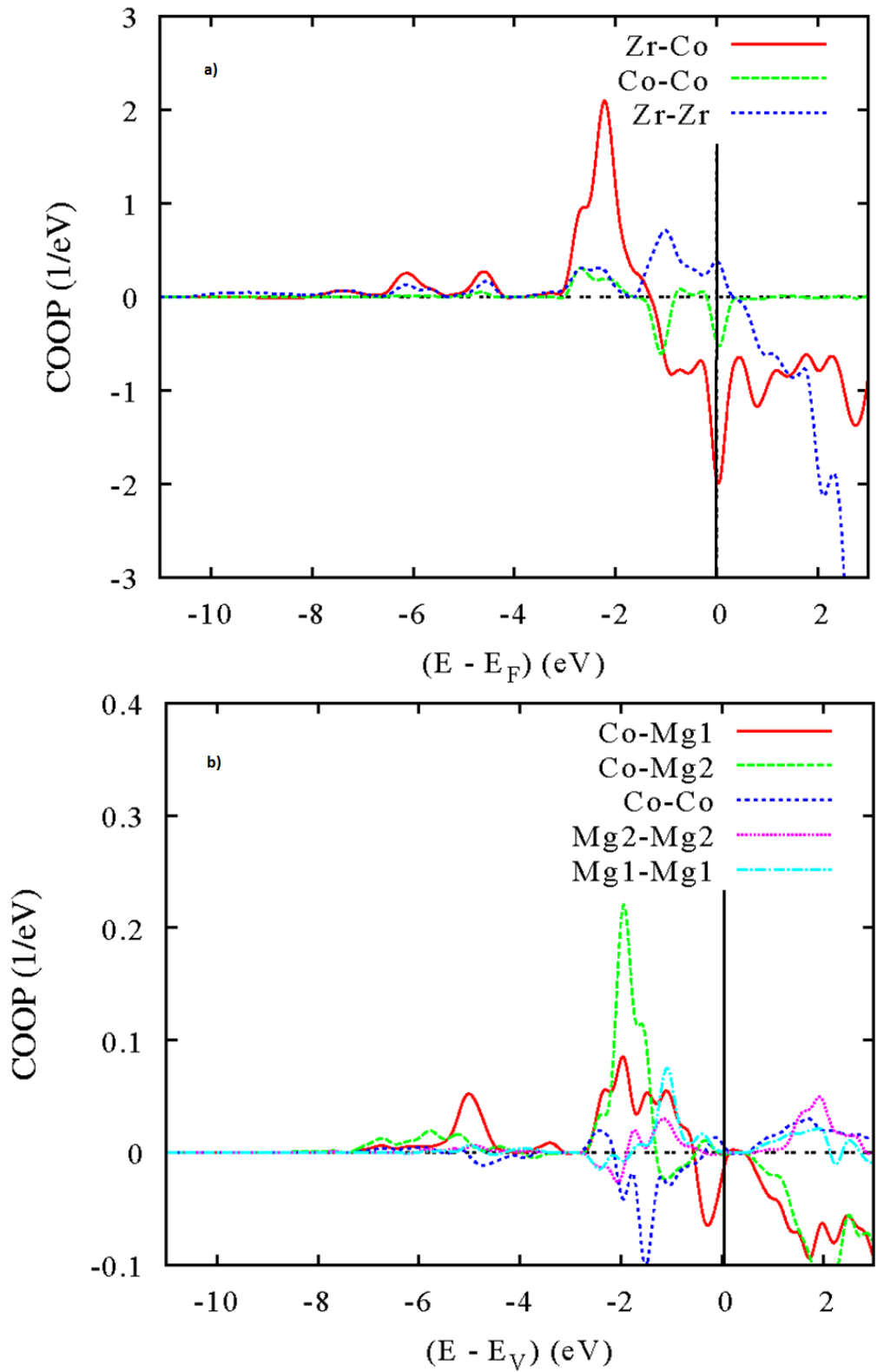

Fig. 4: Chemical bonding for metal-metal pair interactions in a) $Zr_2CoH_5$ and b) $Mg_2CoH_5$.

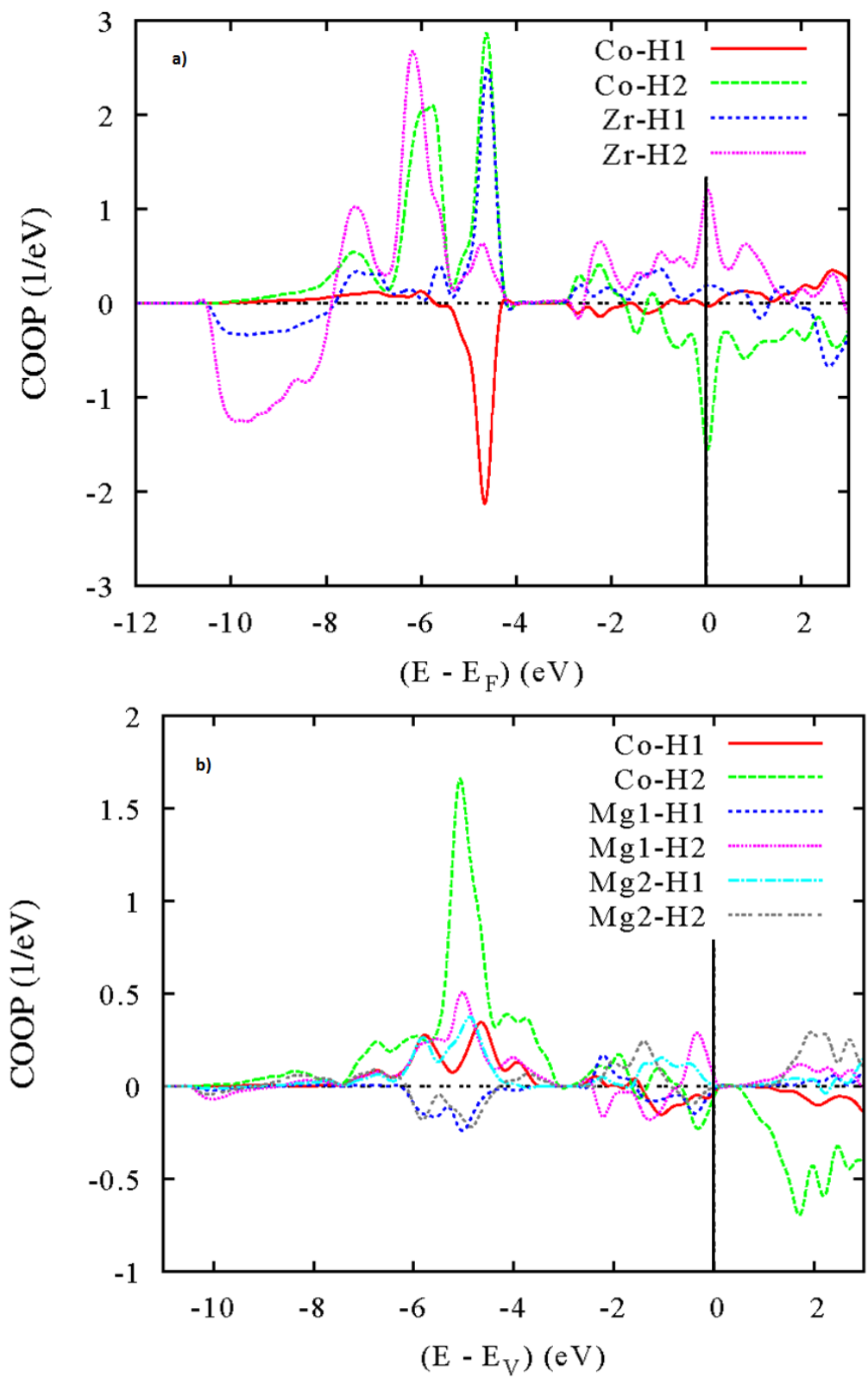

Fig. 5: Chemical bonding for metal-hydrogen pair interactions in a) $Zr_2CoH_5$ and b) $Mg_2CoH_5$.